\documentclass[aps,prd,amsfonts,amsmath,amssymb,twocolumn,10pt,floatfix]{revtex4-1}

\usepackage{hyperref}
\usepackage{graphicx}
\usepackage{color}
\usepackage{multirow}
\usepackage[normalem]{ulem} 

\usepackage[T1]{fontenc}

\hypersetup{
  colorlinks   = true, %Colours links instead of ugly boxes
  urlcolor     = blue, %Colour for external hyperlinks
  linkcolor    = blue, %Colour of internal links
  citecolor   = blue %Colour of citations
}

\newcommand{\dif}{{\rm d}}

\def\apss{Astrophysics and Space Science}
\def\jcap{Journal of Cosmology and Astroparticle Physics}
\def\prd{Physical Review D}

\def\physrep{Physics Reports}
\begin{document}

\title{Vacuum solutions around spherically symmetric and static  objects \\ in the Starobinsky model}

\author{Sercan {\c C}{\i}k{\i}nto{\u g}lu}
\email{cikintoglus@itu.edu.tr}
\affiliation{Istanbul Technical University, Faculty of Science and Letters, Physics Engineering Department, 34469 Maslak, Istanbul, Turkey}

\begin{abstract}
The vacuum solutions around a spherically symmetric
and static object  in the Starobinsky model are studied with a perturbative approach.
The differential equations for the components of the metric and
the Ricci scalar are obtained and solved by using the method of matched asymptotic expansions. 
The presence of higher order terms in this gravity model leads to 
the formation of a boundary layer near the 
surface of the star allowing the accommodation of the extra boundary conditions on the Ricci scalar.
Accordingly, the metric can be different
from the Schwarzschild solution near the star depending on the value of the Ricci scalar
at the surface of the star while matching the Schwarzschild metric far from the star. 
\end{abstract}

\maketitle
\section{Introduction}

A commonly followed path for addressing the accelerated expansion of the Universe is to modify Einstein's general relativity (GR).
One approach to modify GR is to
replace the Einstein-Hilbert Lagrangian with a function of 
the Ricci scalar, the so-called $f\left(\mathcal{R}\right)$ models
\cite{Sot+10,Fel+10,Cap+11,Odin+11}. Among these,
the Starobinsky model \cite{Star80}, $f\left(\mathcal{R}\right)=
\mathcal{R}+\alpha\mathcal{R}^2$, is one of the most popular models of gravity
as it provides a natural inflationary era in the early universe and it does not contain ghostlike modes.

The structures of relativistic stars have been studied to
test $f\left(\mathcal{R}\right)$ models in a strong gravity regime
in Refs.~\cite{Kob+09,Bab+09,
Bab+10,Sal+11,Coo+10,Arap+11,Gang+14,Arap+16,Yaz+14a,Yaz+14b,Yaz+15a,Yaz+15b,
Odin+13,Odin+14,Odin+15a,Odin+15b,Odin+16}. The second term in the Lagrangian 
is assumed to be perturbative in Refs.~\cite{Coo+10,Arap+11}, and the authors reduced the order
by the method of perturbative constraints  \cite{Eli+89,Jae+86}. 
More recently, another perturbative method known as \textit{matched asymptotic expansions} (MAE) \cite{Bend+78,holm12} has been employed \cite{Gang+14,Arap+16} for the singular perturbation problem 
posed by the hydrostatic equations.

The latter approach requires extra boundary conditions since 
the order of the differential equations increases. 
The most reasonable choice for the extra boundary condition
is the value of the Ricci scalar at the surface of the star. Yet,
in $f\left(\mathcal{R}\right)$ theories, the vacuum solutions are not unique
around a spherically symmetric and static object unlike the case in GR.

The vacuum solutions for $f(R)$ theories are studied by employing different approaches
in Refs.~\cite{Mul+06,Cap+07,Pun+08,Gao+16}. 
In these works, 
the authors search for a vacuum solution within $f\left(\mathcal{R}\right)$ models
for specific choices of the Ricci scalar.
Although they show that the
Schwarzschild solution can be a vacuum solution for 
$f\left(\mathcal{R}\right)=\mathcal{R}+\alpha\mathcal{R}^2$,
their results are not unique in the sense that different solutions
of the field equations are also possible.
Besides these works, the exterior solution is found as the
Schwarzschild-de Sitter metric with a perturbative approach
for $f\left(\mathcal{R}\right)$ models where the modified term is a 
power of the Ricci scalar in Ref.~\cite{Coo+10}. A regular perturbative approach where the solutions are expanded as corrections to general relativity's solutions is employed in this work although
the trace equation poses a singular perturbation problem. Accordingly, some solutions might have been missed. 
A different perturbative approach is employed in \cite{Res+16}, and the vacuum solutions are found as a decaying harmonic function of the Ricci scalar
within $\mathcal{R}+\alpha\mathcal{R}^2$ for negative $\alpha$ values. Yet, negative $\alpha$ is known to lead to ``ghosts'' in the theory.

The \textit{no-hair theorem} which states that a black hole is characterized by only its mass, spin, and electric charge
\cite{Isr67,Isr68,Car71}, is known to prevail in modified gravity theories \cite{Hawk72,Mayo+96,Psa+08,Sot+12,Sal+16}. The validity of the no-hair theorem is shown in Ref.\ \cite{Sal+16} for the Starobinsky model with a spherically symmetric and static setup. Also, the authors of Ref.\ \cite{Mig+92} showed that the only static spherically symmetric asymptotically flat solution with a regular horizon is the Schwarzschild solution for positive values of $\alpha$.

In this work, the MAE method, which is appropriate for handling the singular perturbation problem \cite{Bend+78,holm12} posed by the trace equation, is employed to obtain the vacuum solutions around spherically symmetric objects. The results demonstrate the possibility of solutions other than the Schwarzschild solution in the $f\left(\mathcal{R}\right)=\mathcal{R} + \alpha \mathcal{R}^2$ gravity model around relativistic stars. Our results are applied to spherically symmetric and static black holes for vanishing Ricci scalar at the horizon and for this value our solutions reduce to the Schwarzschild metric, in full consistency with these previous results \cite{Mayo+96,Psa+08,Sot+12,Sal+16,Mig+92}.

The plan of the paper is as follows:
The field equations are obtained in
 Sec.\ \ref{sec:FE}. In  Sec.\ \ref{sec:IOS}, these equations are
solved by using the MAE method. Then, the composite solutions are constructed by matching these solutions, and the vacuum solution
is obtained in the Starobinsky model in \ Sec.\ \ref{sec:VS}.
Finally, the vacuum solutions are discussed in  Sec.\ \ref{sec:Conc}.

\section{Field Equations \label{sec:FE}}

The action of the Starobinsky model is
\begin{equation}
S=\frac{1}{16\pi }\int {\rm d}^{4}x\sqrt{-g}\left(\mathcal{R}+\alpha \mathcal{R}^2\right)+S_{\rm matter}, \label{action}
\end{equation}
where $g$ is the determinant of the metric $g_{\mu\nu}$, $\mathcal{R}$ is the Ricci scalar, $\alpha$ is a positive constant, and $S_{\rm matter}$ is the action of matter. 
We assume the second term of the Lagrangian is a perturbative correction to the first term
\cite{Jae+86,Eli+89}. In the metric formalism, the variation of the action with respect to the metric gives the field equations, 
\begin{equation}
\left( 1+2\alpha \mathcal{R}\right) G_{\mu \nu }+\frac{1}{2}\alpha g_{\mu \nu }\mathcal{R}^2
-2\alpha \left( \nabla _{\mu }\nabla _{\nu }-g_{\mu \nu }\square \right)
\mathcal{R}=8\pi T_{\mu \nu }
\end{equation}
\cite{Sot+10,Fel+10}. Contracting with the inverse metric, the trace equation is
\begin{equation}
6\alpha \square \mathcal{R}- \mathcal{R}=8\pi T.
\end{equation}
A general form of the spherically symmetric and static metric is
\begin{equation}
{\rm d}s^{2}=-h\left(r\right){\rm d}t^{2}+f\left(r\right){\rm d}r^{2}+r^{2}\left( {\rm d}\theta
^{2}+\sin ^{2}\theta {\rm d}\phi ^{2}\right).
\end{equation}
All components of the energy-momentum tensor are zero in vacuum.
Accordingly,
 by using the field equations and the trace equation, a set of differential equations for
the metric components and the Ricci scalar in vacuum can be obtained as
\begin{align}
6r&\left( 1+2\alpha \mathcal{R}\right) \left( 1+2\alpha
\mathcal{R}+\alpha r\mathcal{R}^{\prime }\right) {\frac{\dif f }{\mathrm{d}r}} \notag \\
=&
f\left( 1+2\alpha \mathcal{R}\right) 
\left[ 6\left( 1+2\alpha \mathcal{R}\right) \left( 1-f\right)
+{r}^{2}f\mathcal{R}\left( 2+3\alpha\mathcal{R}\right) \right]   \notag
\\
& +2\alpha f
\left[ 6r\left( 1+2{\alpha}\mathcal{R}\right)\left( 1-f\right) 
+r^{3}f\mathcal{R}\left( 1+3\alpha \mathcal{R}\right) \right] \mathcal{R}^{\prime }
\notag \\
& +24{\alpha}^{2}f{r}^{2}\mathcal{R}^{\prime 2}, \label{diff_f} \\
-&\left( 1+2\alpha \mathcal{R}+\alpha r\mathcal{R}^{\prime }\right)
\frac{\dif h}{\mathrm{d}r} \notag \\
&=
\frac{h}{r}
\left[ 1-f
+\alpha \left( 
\frac{r^2f\mathcal{R}^2}{2}+2\left(1-f\right)\mathcal{R}+4r\mathcal{R}^{\prime }
\right) \right], \label{diff_h} \\
6\alpha&\left( 1+2\alpha\mathcal{R}\right)r\frac{\dif^2\mathcal{R}}{\dif r^2} 
\notag \\
=&
rf\mathcal{R}\left( 1+2\alpha\mathcal{R}\right)  
-\alpha\mathcal{R}^{\prime }\left[ 
6\left( 1+2\alpha \mathcal{R}\right) \left( 1+f\right)
\right.
\notag \\ 
&\left.
-\left( 1+3\alpha\mathcal{R}\right) r^{2}f\mathcal{R}
\right]  
+12{\alpha}^{2}r\mathcal{R}^{\prime 2}. \label{diffeq_R}
\end{align}%
Considering the third equation, this set of differential equations poses a
singular perturbation problem \cite{Bend+78,holm12},
since $\alpha$ is a small parameter.
To solve these equations,
we will use the MAE method which is appropriate for such problems.
According to the MAE method, a \textit{boundary layer} occurs
where the highest derivative term is non-negligible compared with the other terms.
The location of the boundary layer is not known from the beginning. Moreover, there could be more than one boundary layer. Instead of trying to
locate the boundary layer, we work on a fictitious finite size domain where the
boundary layer occurs at one of the edges of the region. 
Later, we will check which choices give physical solutions and construct the vacuum solutions   accordingly.
We use $R_-$ and $R_+$ to denote the nearest and the farthest points of the region to the star, respectively.
To employ the MAE method, parameters should be nondimensionalized. With the definitions
\begin{equation}
x \equiv \frac{r}{L}-L_{\ast}, \qquad L\equiv  R_+-R_-, \qquad L_{\ast} \equiv \frac{R_-}{L},
\label{nondimx}
\end{equation}
the problem is restricted to the interval $0<x<1$, and this leads to
\begin{equation}
r=\left(x+L_{\ast}\right)L, \qquad
\frac{\dif}{\dif r}=\frac{1}{L}\frac{\dif}{\dif x}, \qquad
\frac{\dif^2}{\dif r^2}=\frac{1}{L^2}\frac{\dif^2}{\dif x^2}.
\end{equation}
The other parameters can be made dimensionless by using finite scale factor $L$ as
\begin{equation}
\mathcal{\bar{R}}=L^2 \mathcal{R}, \qquad 
\epsilon = \frac{\alpha}{L^2}. \label{nondim2}
\end{equation}
Then Eqs. \eqref{diff_f}, \eqref{diff_h}, and
\eqref{diffeq_R}, respectively, become
\begin{align}
\big( 1+&2\epsilon\mathcal{\bar{R}}
+\epsilon\left(x+L_{\ast}\right)\mathcal{\bar{R}}^{\prime }
\big) 
\frac{\mathrm{d}f}{\mathrm{d}x}
\notag \\
=& 
\frac{f}{6} 
\left[ 6\left(
1+2\epsilon\mathcal{\bar{R}}\right) 
\frac{1-f}{x+L_{\ast}}
+{\left(x+L_{\ast}\right)}f \mathcal{\bar{R}}
\left( 2+3\epsilon \mathcal{\bar{R}}\right) %
\right]   \notag \\
& +2\epsilon f\left[
1-f
+
\left(x+L_{\ast}\right)^{2}\frac{f}{6}
\frac{\left(1+3\epsilon\mathcal{\bar{R}}\right)}
{\left(1+2\epsilon\mathcal{\bar{R}}\right)}
\mathcal{\bar{R}}
\right] 
\mathcal{\bar{R}}^{\prime }  \notag \\
& +4\epsilon^{2}\left(x+L_{\ast}\right)
\frac{f}{1+2\epsilon\mathcal{\bar{R}}}
\mathcal{\bar{R}}^{\prime 2},  
\label{diff_dimf} \\
\big( 1+&2\epsilon\mathcal{\bar{R}}+\epsilon\left(x+L_{\ast}\right)\mathcal{\bar{R}}^\prime\big)
\frac{\dif h}{\dif x}
\notag \\
=&
-\frac{h}{x+L_{\ast} } 
\left[ 1-f+\epsilon \left( \frac{\left(x+L_{\ast}\right)^{2}f\mathcal{\bar{R}}^{2}}{2}\right.\right.
\notag \\
&+2\left(1-f\right) \mathcal{\bar{R}}
+4\left(x+L_{\ast}\right)\mathcal{\bar{R}}^{\prime }
\Bigg) \Bigg], 
\label{diff_dimh} \\
6\epsilon \big( 1&+2\epsilon\mathcal{\bar{R}}\big) \left(x+L_{\ast}\right)
\frac{\mathrm{d}^{2}\mathcal{\bar{R}}}{\mathrm{d}{x}^{2}} 
\notag \\
=&
\left(x+L_{\ast}\right)f\mathcal{\bar{R}}\left( 1+2\epsilon\mathcal{\bar{R}}\right)  
-\epsilon \mathcal{\bar{R}}^{\prime }\left[ 6\left( 1+2\epsilon\mathcal{\bar{R}}\right) \left( 1+f\right) \right.
\notag \\
&\left.-\left( 1+3\epsilon\mathcal{\bar{R}}\right) 
\left(x+L_{\ast}\right)^{2}f\mathcal{\bar{R}}\right]
+12\epsilon^2\left(x+L_{\ast}\right)\mathcal{\bar{R}}^{\prime 2}. \label{diff_dimR}
\end{align}%

Far from the object the metric should converge to the Minkowski metric. Therefore, the most reasonable boundary 
conditions for the components of the metric are $f\left( r\rightarrow
\infty \right) =1$ and $h\left( r\rightarrow \infty \right) =1$, and
for the Ricci scalar it is $\mathcal{R}\left( r\rightarrow \infty \right) =0$.
The Starobinsky model is an exceptional case, among $f(\mathcal{R})$ theories, in that the Ricci scalar can be discontinuous at the surface of the object in the presence of thin shells or braneworlds \cite{Sen13}. As we assume absences of braneworlds and thin shells in this paper,  the interior of the star continuously matches with the exterior without any need for the Chameleon mechanism. This requires the continuity of the Ricci scalar and its derivative on the surface of the star according to junction conditions derived in \cite{Sen13}. So, we choose the final boundary condition as 
$\mathcal{R}\left( r=R_{\ast} \right)=\mathcal{R}_{\ast}$
where $R_{\ast}$ is the radius of the star and $\mathcal{R}_{\ast}$ is the Ricci scalar at the
surface of the star which is provided by the interior solutions of the star. The condition of continuity of Ricci scalar's derivative can be used to test the validity of the solutions.

\section{Inner and Outer Solutions \label{sec:IOS}}

According to the method, we need to seek solutions inside and outside the boundary layer
separately.
The solutions valid inside the boundary layer
are called the \textit{inner solutions}, and the solutions
valid outside the boundary layer
are called the \textit{outer solutions}. Both of the solutions are introduced as perturbative series. The \textit{composite solutions}, which are valid 
all over the interval of the problem, are constructed
by combining the inner and outer solutions after matching them \cite{Bend+78,holm12}.

\subsection{Outer solutions}

We can obtain the general outer solutions 
before deciding the location of the boundary layer. 
By introducing the outer solution of the Ricci scalar as a perturbative series,
\begin{equation}
\mathcal{\bar{R}}^{\rm out}\left(x\right) =
\mathcal{\bar{R}}^{\rm out}_0\left(x\right)
+\epsilon \mathcal{\bar{R}}^{\rm out}_1\left(x\right)
+O\left(\epsilon^2\right),
\end{equation}
Eq.\ \eqref{diff_dimR} can be written up to the first order as 
\begin{align}
6\epsilon &\left(x+L_{\ast}\right){\frac{\mathrm{d}^{2}\mathcal{\bar{R}}^{\rm out}_0}
{\mathrm{d}{x}^{2}}} 
\notag \\
=&\left(x+L_{\ast}\right)f^{\rm out}\mathcal{\bar{R}}^{\rm out}_0
+\epsilon \left(x+L_{\ast}\right)f^{\rm out} 
\left[
\mathcal{\bar{R}}^{\rm out}_1
+2 \left(\mathcal{\bar{R}}^{{\rm out}}_0\right)^2
\right] \notag \\
&-\epsilon
\mathcal{\bar{R}}^{\rm out \prime}_0
\left( 6+6f^{\rm out} -
\left(x+L_{\ast}\right)^{2}f^{\rm out}
\mathcal{\bar{R}}^{\rm out}_0 
\right).
\end{align}%
By solving the equation order by order, the solutions, 
independent from the boundary conditions, are obtained as
\begin{equation}
\mathcal{\bar{R}}^{\rm out}_0 = \mathcal{\bar{R}}^{\rm out}_1 = 0, \label{out_sol_R}
\end{equation}
since $f^{\rm out}\left(x\right)=0$ is not physical.
Therefore, Eqs.\ \eqref{diff_dimf} and \eqref{diff_dimh} become 
\begin{align}
\frac{\mathrm{d}f^{\rm out}}{\mathrm{d}x}=& 
\frac{f^{\rm out}}{x+L_{\ast}}\left( 1-f^{\rm out} \right),   \\ 
\frac{\mathrm{d}h^{\rm out}}{\mathrm{d}x}=&
-\frac{h^{\rm out}}{x+L_{\ast}}\left(1-f^{\rm out}\right).
\end{align}%
These equations do not contain a perturbative part and 
they are similar to the GR case.
The general solutions of these equations are
\begin{align}
f^{\rm out}\left(x\right) =& \frac{L_{\ast}+x}{A+x}, \label{out_sol_f}\\
h^{\rm out}\left(x\right) =& B\frac{A+x}{L_{\ast}+x}. \label{out_sol_h}
\end{align}

\subsection{Inner solutions}

We do not have any mathematical justification for the
location of the boundary layer. Yet, the Ricci scalar is
zero outside the boundary layer as found in the previous section. Then, a boundary layer
which occurs at the farthest point of the fictitious region 
requires deviation of the Ricci scalar from zero 
with the divergence of the metric components from one (see Appendix \ref{app:C2}).
That corresponds to a solution in which gravity increases radially away from the source in the vacuum which is not physical. So, 
a boundary layer can occur only at the nearest point of the fictitious region.

In a regular perturbative approach,
due to the factor of $\epsilon$, the highest order differential term does not appear in the differential equations,
and some solutions are missed because of this order reduction. According to the MAE method, the
boundary layer is where the highest order differential term becomes non-negligible,
and  
to appropriately examine solutions inside the boundary layer 
we need to define a new coordinate variable (coordinate stretching parameter)
\begin{equation}
\xi \equiv \frac{x}{\epsilon^n}.
\end{equation}
With this definition, the derivative, $\dif/\dif x$, reduces the order of the term upon 
which it acts, as $n$. Hence, the highest order differential term becomes more significant.

\begin{widetext}
In terms of the inner variable, Eq.\ \eqref{diff_dimR} turns into
\begin{align}
6\left( 1+2\epsilon\mathcal{\bar{R}}^{\rm in} \right)
\left( \epsilon^{1-n}\xi+\epsilon^{1-2n}L_{\ast} \right)
\frac {\dif^{2}\mathcal{\bar{R}}^{\rm in}}{\dif\xi^2} =& 
\left( \epsilon^n\xi+L_{\ast} \right) 
\left( 1+2\epsilon\mathcal{\bar{R}}^{\rm in} \right)
f^{\rm in}\mathcal{\bar{R}}^{\rm in} 
-\left[
6\epsilon^{1-n}\left(1+f^{\rm in}\right) 
+12\epsilon^{2-n}\left( 1+f^{\rm in}\right)
\mathcal{\bar{R}}^{\rm in}
\right]
\left(\mathcal{\bar{R}}^{\rm in}\right)^\prime
\notag \\
&
+f^{\rm in}\left[ 
\epsilon^{1+n}\xi^{2}\mathcal{\bar{R}}^{\rm in}
+2\epsilon\xi L_{\ast}\mathcal{\bar{R}}^{\rm in}
+
{\epsilon}^{1-n}L_{\ast}^2\mathcal{\bar{R}}^{\rm in} 
+3\epsilon^{2+n}{\xi}^{2}
\left(\mathcal{\bar{R}}^{\rm in}\right)^{2}
\right]
\left(\mathcal{\bar{R}}^{\rm in}\right)^\prime 
\notag \\
&
+f^{\rm in}\left[
6\epsilon^2\xi L_{\ast}\left(\mathcal{\bar{R}}^{\rm in}\right)^{2}
+3\epsilon^{2-n}L_{\ast}^2\left(\mathcal{\bar{R}}^{\rm in}\right)^{2}
\right]
\left(\mathcal{\bar{R}}^{\rm in}\right)^\prime 
\notag \\
&+12\left(\epsilon^{2-n}\xi+\epsilon^{2-2n}L_{\ast} \right)
\left( \mathcal{\bar{R}}^{\rm in}\right)^{\prime 2}.
\label{in_dif_R_c}
\end{align} 
$n$ should be chosen such that the highest order differential term and one of the other terms become the lowest order terms in the equation. Hence, the order reduction does not occur and
no solutions are missed.
Accordingly, $n=1/2$ is the most suitable choice to balance
the second order differential term in the left-hand side of the equation
with another term in the equation.
Hence, the above equation becomes
\begin{align}
6(1+2\epsilon\mathcal{\bar{R}}^{\rm in} ) 
( \sqrt{\epsilon}\xi+L_{\ast} )
\frac{\dif^2\mathcal{\bar{R}}^{\rm in}}{\dif\xi^2} =&
\left( \sqrt{\epsilon}\xi+L_{\ast} \right)  
\left( 1+2\epsilon\mathcal{\bar{R}}^{\rm in}\right)
f^{\rm in}\mathcal{\bar{R}}^{\rm in}
-6\sqrt{\epsilon}
\left(1+f^{\rm in} \right)
\left( 1+2\epsilon\mathcal{\bar{R}}^{\rm in} \right)
\left(\mathcal{\bar{R}}^{\rm in}\right)^\prime 
\notag \\
&
+(\epsilon^{3/2}\xi^2
+2\epsilon\xi L_{\ast}
+\sqrt{\epsilon}L_{\ast}^2)
( 1+3\epsilon\mathcal{\bar{R}}^{\rm in})
f^{\rm in}\mathcal{\bar{R}}^{\rm in}
(\mathcal{\bar{R}}^{\rm in} )^\prime  
+12( \epsilon^{3/2}\xi+\epsilon L_{\ast} ) 
( \mathcal{\bar{R}}^{\rm in} )^{\prime 2} . \label{in_dif_R_c1}
\end{align}
Similarly, Eqs.\ \eqref{diff_dimf} and \eqref{diff_dimh} become
\begin{align}  
\left[ 1+2\epsilon\mathcal{\bar{R}}^{\rm in} 
+\sqrt{\epsilon} \left( \xi\sqrt{\epsilon}+L_{\ast} \right) 
\left(\mathcal{\bar{R}}^{\rm in}\right)^\prime
\right]
{\frac{\dif f^{\rm in}}{\dif\xi}}=& 
\frac{\sqrt{\epsilon}f^{\rm in}}
{\xi\sqrt{\epsilon}+L_{\ast}}  
\left( 1+2\epsilon\mathcal{\bar{R}}^{\rm in}\right)  
\left( 1-f^{\rm in} \right)  
+\frac{\sqrt{\epsilon}}{6}
f^{\rm in}  
\left( \xi\sqrt{\epsilon}+L_{\ast} \right)
\left( 2+3\epsilon \mathcal{\bar{R}}^{\rm in} \right) 
f^{\rm in}\mathcal{\bar{R}}^{\rm in}
\notag \\
&
+2\epsilon f^{\rm in}\left(\mathcal{\bar{R}}^{\rm in}\right)^\prime
\left(1-f^{\rm in}\right) 
+\epsilon \frac{\left(f^{\rm in}\right)^2
\left(\mathcal{\bar{R}}^{\rm in}\right)^\prime}
{3\left( 1+2\epsilon\mathcal{\bar{R}}^{\rm in} \right)}
\left( \xi\sqrt{\epsilon}+L_{\ast} \right)^2
\left( 1+3\epsilon\mathcal{\bar{R}}^{\rm in}\right)
\mathcal{\bar{R}}^{\rm in} 
\notag \\
&
+4\epsilon^{3/2}f^{\rm in}
\left( \xi\sqrt{\epsilon}+L_{\ast}\right) 
\frac{\left( \mathcal{\bar{R}}^{\rm in}\right) ^{\prime 2}}
{\left( 1+2\epsilon\mathcal{\bar{R}}^{\rm in} \right)}
, \label{in_dif_f_c1}
\end{align}
and
\begin{align}
\frac{\dif h^{\rm in}}{\dif\xi}=&
-\frac{h^{\rm in}}
{\left( \xi\sqrt{\epsilon}+L_{\ast} \right) 
\left( 1+2\epsilon\mathcal{\bar{R}}^{\rm in}
+\sqrt{\epsilon}\left( \xi\sqrt{\epsilon}+L_{\ast} \right) 
\left(\mathcal{\bar{R}}^{\rm in} \right)^\prime\right)} 
\label{in_dif_h_c1} \\
&
\left\lbrace 
\sqrt{\epsilon}\left(1-f^{\rm in}\right)
+
\epsilon \left[ 
\frac{\sqrt{\epsilon}}{2}
\left( \xi\sqrt{\epsilon}+L_{\ast}\right)^{2}f^{\rm in}
\left( \mathcal{\bar{R}}^{\rm in}\right)^{2}
+2\sqrt{\epsilon}\left(1-f^{\rm in}\right) 
\mathcal{\bar{R}}^{\rm in}
+
4\left( \xi\,\sqrt{\epsilon}+L_{\ast} \right)
\left(\mathcal{\bar{R}}^{\rm in}\right)^\prime
 \right] \right\rbrace .
\notag 
\end{align}
The inner solutions are introduced as perturbed series,
\begin{align}
f^{\rm in}\left(\xi\right)=&f^{\rm in}_0\left(\xi\right) 
+\sqrt{\epsilon}f^{\rm in}_1\left(\xi\right)
+\epsilon f^{\rm in}_2\left(\xi\right) 
+O\left(\epsilon^{3/2}\right),
\\
h^{\rm in}\left(\xi\right)=&h^{\rm in}_0\left(\xi\right) 
+\sqrt{\epsilon}h^{\rm in}_1\left(\xi\right)
+\epsilon h^{\rm in}_2\left(\xi\right)
+O\left(\epsilon^{3/2}\right),
\\
\mathcal{R}^{\rm in}\left(\xi\right)=&
\mathcal{R}^{\rm in}_0\left(\xi\right) 
+\sqrt{\alpha}\mathcal{R}^{\rm in}_1\left(\xi\right)
+\epsilon \mathcal{R}^{\rm in}_2\left(\xi\right)
+O\left(\epsilon^{3/2}\right).
\end{align}
After plugging these solutions into Eqs.\
\eqref{in_dif_f_c1}, \eqref{in_dif_h_c1}, \eqref{in_dif_R_c1}, 
they are found as 
\begin{align}
f_0^{\rm in}\left(\xi\right) =& C_0, \\
f_1^{\rm in} \left( \xi \right) =&
-\sqrt{\frac{2C_0^3}{3}}H_0L_{\ast}
\exp\left(-\sqrt{\frac{C_0}{6}}\xi\right)
+C_0\frac{1-C_0}{L_{\ast}}\xi +C_1, \\
f_2^{\rm in} \left( \xi \right) =&
\left[
\frac{H_0C_0^2}{12}\left(1-C_0\right)\xi^2
-\frac{7\sqrt{6}C_0^{3/2}H_0}{12}\left(1-C_0\right)\xi
+\frac{L_{\ast}}{6}C_0C_1H_0\xi
\right]
\exp\left(-\sqrt{\frac{C_0}{6}}\xi\right)
\notag \\
&
+\left[
\frac{C_0H_0}{4}\left(1-5C_0\right)
-\frac{5L_{\ast}C_1H_0}{12}\sqrt{6C_0}
-\frac{1}{3}L_{\ast}\sqrt {6}C_0^{3/2}H_1
\right]
\exp\left(-\sqrt{\frac{C_0}{6}}\xi\right)
\notag \\
&
+\frac{5}{6}L_{\ast}^2H_0^2C_0^2
\exp\left(-2\sqrt{\frac{C_0}{6}}\xi\right)
+\frac{C_0^2}{L_{\ast}^2}\left(C_0-1\right)\xi^2
+\frac{C_1}{L_{\ast}}\left(1-2C_0\right)\xi
+C_2,
\\
h_0^{\rm in}\left(\xi\right) =& D_0, \\
h_1^{\rm in}\left(\xi\right) =& D_0\frac{C_0-1}{L_{\ast}}\xi+D_1, \\
h_2^{\rm in} \left( \xi \right) =&
\left(C_0-3\right)D_0H_0\exp\left(-\sqrt{\frac{C_0}{6}}\xi\right)
+\frac{D_1\left(C_0-1\right)+C_1D_0}{L_{\ast}}\xi
+\frac{D_0}{L_{\ast}^2}\left(1-C_0\right)\xi^2
+D_2,
\end{align}
\begin{align}
\mathcal{\bar{R}}_0^{\rm in}\left(\xi\right)=&
H_0\exp\left(-\sqrt{\frac{C_0}{6}}\xi\right), \\
\mathcal{\bar{R}}_1^{\rm in}\left(\xi\right)=&
H_1\exp\left(-\sqrt{\frac{C_0}{6}}\xi\right) 
-\frac{H_0}{24L_{\ast}C_0}
\left[ 
\sqrt{6C_0^3}\left(1-C_0\right)\xi^2
+\left(6C_0^2+18C_0+2L_{\ast}C_1\sqrt{6C_0}\right)\xi
\right.
\notag \\
&\left.
+6L_{\ast}C_1+3\sqrt{6C_0^3}+9\sqrt{6C_0}
\right] 
\exp\left(-\sqrt{\frac{C_0}{6}}\xi\right)
-L_{\ast}H_0^2\sqrt{\frac{C_0}{6}}
\exp\left(-2\sqrt{\frac{C_0}{6}}\xi\right),
\\
\mathcal{\bar{R}}_2^{\rm in}\left(\xi\right)=&
H_2\exp\left(-\sqrt{\frac{C_0}{6}}\xi\right) 
+\frac{H_0^3L_{\ast}^2C_0}{4}
\exp\left(-3\sqrt{\frac{C_0}{6}}\xi\right)
+C_1H_0
\frac{\sqrt{6C_0}\left(21C_0+15\right)
+6C_1L_{\ast}}
{288L_{\ast}C_0}
\xi^2
\exp\left(-\sqrt{\frac{C_0}{6}}\xi\right)
\notag \\
&
+\frac{H_0^2}{12}
\left[ 
-C_0\left(C_0-1\right)\xi^2
+2\left( C_1L_{\ast}+\sqrt{6C_0^3}\right)\xi
-\frac{4H_1}{H_0}L_{\ast}\sqrt{6C_0} 
\right]
\exp\left(-2\sqrt{\frac{C_0}{6}}\xi\right)
\notag \\
&
+\left[ 
\frac{C_0H_0}{192L_{\ast}^2}\left( C_0-1 \right)^2\xi^4
-H_0\sqrt{6}
\frac{C_0^{1/2}\left(9C_0+11\right)+C_1L_{\ast}\sqrt{6}}
{288L_{\ast}^2}
\left(C_0-1\right)\xi^3
\right]
\exp\left(-\sqrt{\frac{C_0}{6}}\xi\right)
\notag \\
&
+
\frac{
36C_0H_0\left(C_0+1\right)
+12H_1L_{\ast}\sqrt{6C_0}\left(C_0-1\right)
+216H_0
}
{288L_{\ast}^2}
{\xi}^{2}
\exp\left(-\sqrt{\frac{C_0}{6}}\xi\right)
\notag \\
&
-\sqrt{6}C_1
\frac{4L_{\ast}C_0H_1
+H_0\sqrt{6C_0}\left(C_0-3\right)
-2L_{\ast}C_1H_0
}
{48C_0^{3/2}L_{\ast}}\xi
\exp\left(-\sqrt{\frac{C_0}{6}}\xi\right)
\notag \\
&
-\frac{ 
2\sqrt{6C_0}L_{\ast}H_1\left(C_0+3\right)
-3C_0^2H_0
+4L_{\ast}^2H_0C_2+6H_0C_0
-27H_0
}
{8L_{\ast}^2\sqrt{6C_0}}\xi
\exp\left(-\sqrt{\frac{C_0}{6}}\xi\right).
\end{align}
\end{widetext}
Here, by taking some arbitrary constants zero, positive powers of $\exp\left(\xi\right)$ are removed to prevent infinities in the matching procedure of the inner solutions with the outer solutions.
If the boundary condition $\mathcal{\bar{R}}^{\rm in}\left(0\right)=0$ is employed,
the remaining exponential terms would also vanish. So, with this choice of the boundary condition, a boundary layer does not
occur and the vacuum solutions would be the same 
as Schwarzschild's solution.

\section{Vacuum Solutions \label{sec:VS}}

The outer solutions are valid outside the boundary layer and the inner solutions are valid inside the boundary layer. These two solutions should converge to each other near the edge of the boundary layer.
So, the asymptotic values of the inner solutions should equal the outer solutions for the small values of  $x$. We will use Van Dyke's method \cite{dyke64} to satisfy this condition. According to the method,
the inner solutions are written in terms of $x$, and then they are expanded up to $O\left(\epsilon\right)$ 
in the limit of $\epsilon\rightarrow 0$. After that, they are equalized to the outer solutions
and the matching conditions are obtained as
\begin{equation}
A=\frac{L_{\ast}}{C_0}, \qquad B=C_0D_0, 
\qquad C_1=C_2=D_1=D_2=0.
\end{equation}

 If a boundary layer occurs at the nearest point of a fictitious region except for the surface of the star, another boundary layer should also occur at the farthest point of the neighboring region simultaneously. 
Since a boundary layer at the farthest point of a fictitious region is not 
reasonable, as mentioned earlier, the boundary layer
can occur only at the surface of the star. 
The outer solutions are dominant outside the boundary layer, and
these solutions are valid in the rest of the vacuum
since there is not another boundary layer. 
So, the outer solutions should satisfy the boundary conditions
$f\left(r\rightarrow\infty\right)=1$  and
$h\left(r\rightarrow\infty\right)=1$ which imply
$C_0D_0 =1$.

The composite solutions are constructed by adding the inner solutions to the outer solutions and
then subtracting the overlapping part.
We can write the composite solutions in dimensional form
by using the definitions given in Eqs.\ \eqref{nondimx}
and \eqref{nondim2}. Here, $R_-$ equals the radius of
the star, $R_{\ast}$, since the boundary layer can occur only at
the surface of the star. Also, the metric should be consistent with Newtonian gravity 
at the weak field limit. The value of the Ricci scalar can be
assumed negligible at the surface of a nonrelativistic
object. Equaling the ``tt'' component
of the metric to $1-2M$ in this limit gives $C_0=\left(1-2M/R_{\ast}\right)^{-1}$
where $G=c=1$. Hence, the dimensional vacuum solutions are
\begin{widetext}
\begin{align}
f^{\rm comp}\left(r\right) =& 
\left( 1 - \frac{2M}{r} \right)^{-1}
-\sqrt{\frac{2C_0^3\alpha}{3}}K_0R_*
\exp\left[-\sqrt{\frac{C_0}{6\alpha}}\left(r-R_*\right) \right]
\notag \\
&
+\left(1-C_0\right)\frac{K_0}{12}\left[
C_0^2\left(r-R_*\right)^2
-7\sqrt{6C_0^3\alpha}\left(r-R_*\right)
\right]
\exp\left[-\sqrt{\frac{C_0}{6\alpha}}\left(r-R_*\right) \right]
\notag \\
&
+\alpha\left[
\frac{C_0K_0}{4}(1-5C_0)
-\frac{\sqrt{6}}{3}R_*C_0^{3/2}K_1
\right]
\exp\left[-\sqrt{\frac{C_0}{6\alpha}}(r-R_*) \right]
\notag \\
&+\alpha\frac{5}{6}(R_*K_0C_0)^2
\exp\left[-2\sqrt{\frac{C_0}{6\alpha}}(r-R_*) \right],
\label{f_vac}
\\
h^{\rm comp}\left(r\right) =& 
1 - \frac{2M}{r}
+\alpha\left(C_0-3\right)\frac{K_0}{C_0}
\exp\left[-\sqrt{\frac{C_0}{6\alpha}}\left(r-R_*\right) \right],
\label{h_vac} \\
\mathcal{R}^{\rm comp}\left(r\right)=&
K_0\exp\left[-\sqrt{\frac{C_0}{6\alpha}}\left(r-R_*\right) \right]
+\sqrt{\alpha}K_1
\exp\left[-\sqrt{\frac{C_0}{6\alpha}}\left(r-R_*\right) \right]
\notag \\
&
+
\frac{
K_1R_*\sqrt{6C_0}\left(C_0-1\right)
+3K_0\left(C_0^2+C_0+6\right)
}
{24R_*^2}
\left(r-R_*\right)^2
\exp\left[-\sqrt{\frac{C_0}{6\alpha}}\left(r-R_*\right)\right]
\notag \\
&
-\sqrt{\alpha}
\frac{ 
2\sqrt{6C_0}R_*K_1\left(C_0+3\right)
-3\left(C_0^2-2C_0+9\right)K_0
}
{8R_*^2\sqrt{6C_0}}\left(r-R_*\right)
\exp\left[-\sqrt{\frac{C_0}{6\alpha}}\left(r-R_*\right)\right]
\notag \\
&
-\sqrt{\alpha}\frac{K_0}{24R_*}
\left[ 
\sqrt{6C_0}\left(1-C_0\right)
\frac{\left(r-R_*\right)^2}{\alpha}
+\left(6C_0+18\right)\frac{r-R_*}{\sqrt{\alpha}}
+3\sqrt{6C_0}+9\sqrt{\frac{6}{C_0}}
\right] 
\exp\left[-\sqrt{\frac{C_0}{6\alpha}}\left(r-R_*\right) \right]
\notag \\
&
-\sqrt{\frac{\alpha C_0}{6}}
R_*K_0^2\exp\left[-2\sqrt{\frac{C_0}{6\alpha}}\left(r-R_*\right) \right] 
+\alpha K_2
\exp\left[-\sqrt{\frac{C_0}{6\alpha}}\left(r-R_*\right)\right] 
+\alpha\frac{K_0^3R_*^2C_0}{4}
\exp\left[-3\sqrt{\frac{C_0}{6\alpha}}\left(r-R_*\right)\right]
\notag \\
&
+\alpha\frac{K_0^2}{12}
\left[ 
-C_0\left(C_0-1\right)\frac{\left(r-R_*\right)^2}{\alpha}
+2\sqrt{6C_0^3}
\frac{r-R_*}{\sqrt{\alpha}}
-\frac{4K_1}{K_0}R_*\sqrt{6C_0} 
\right]
\exp\left[-2\sqrt{\frac{C_0}{6\alpha}}\left(r-R_*\right)\right]
\notag \\
&
+\alpha \frac{K_0}{R_*^2}\left[ 
\frac{C_0}{192}\left( C_0-1 \right)^2
\frac{\left(r-R_*\right)^4}{\alpha^2}
-\sqrt{6C_0}
\frac{\left(9C_0+11\right)}{288}
\left(C_0-1\right)
\frac{\left(r-R_*\right)^3}{\alpha^{3/2}}
\right]
\exp\left[-\sqrt{\frac{C_0}{6\alpha}}\left(r-R_*\right)\right],
\label{R_vac}
\end{align}
where
\begin{equation}
K_0 = \frac{H_0}{L^2}, \qquad K_1 = \frac{H_1}{L^3},
\qquad K_2 = \frac{H_2}{L^4}.
\end{equation}

Accordingly, the nonexponential terms of
the metric components are the same as Schwarzschild's metric. 
The metric converges to the Schwarzschild metric far from the star since the exponential terms go to zero rapidly when $r-R_{\ast}$ becomes 
much greater than $\sqrt{\alpha}$.

So far, we have used three boundary conditions and the Newtonian limit of the metric. By using these, we determined all arbitrary constants except $K_0$, $K_1$, and $K_2$. We  still have freedom to use one more boundary condition to determine these three constants.
The continuity of the Ricci scalar on the surface of the star is sufficient to determine $K_0$, $K_1$, and $K_2$. Still, the continuity of the derivative of the Ricci scalar is required according to the junction conditions derived in \cite{Sen13}. Then, the solutions  which are obtained by solving the interior of the star should satisfy
\begin{align}
\left.\frac{\dif \mathcal{R}^{\rm interior}}{\dif r}\right|_{r=R_*}=&
-K_0\sqrt{\frac{C_0}{6\alpha}}
-K_1\sqrt{\frac{C_0}{6}} 
-\frac{K_0}{24R_*}
\left( 
3C_0+9
\right) 
+\frac{C_0}{3}R_*K_0^2
-\sqrt{\alpha}
\frac{ 
2\sqrt{6C_0}R_*K_1(C_0+3)
-3(C_0^2-2C_0+9)K_0
}
{8R_*^2\sqrt{6C_0}}
\notag \\
&
-\sqrt{\alpha\frac{C_0}{6}}K_2
-3\sqrt{\alpha\frac{C_0}{6}}\frac{K_0^3R_*^2C_0}{4}
+\sqrt{\alpha}\frac{K_0^2}{12}
\left[ 
2\sqrt{6C_0^3}+8\frac{K_1}{K_0}R_*C_0
\right]. \label{Rprim_con}
\end{align}
\end{widetext}

\begin{figure*}[h!]
\center
\includegraphics{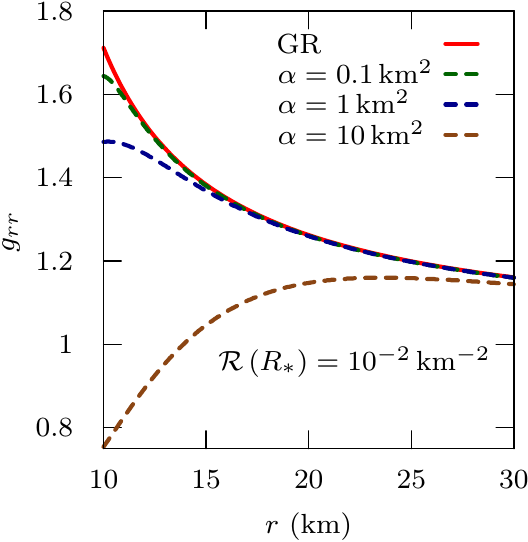}
\includegraphics{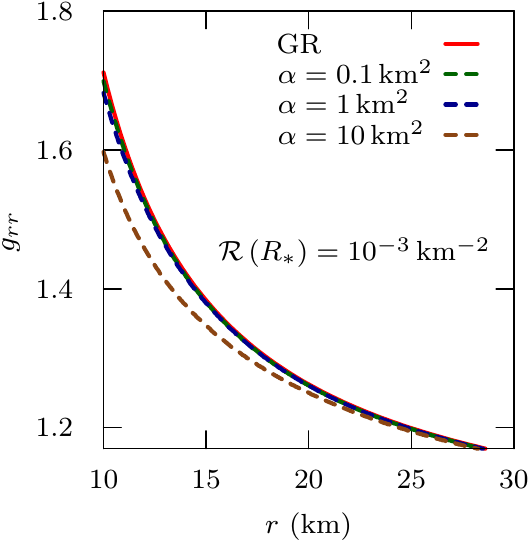}
\includegraphics{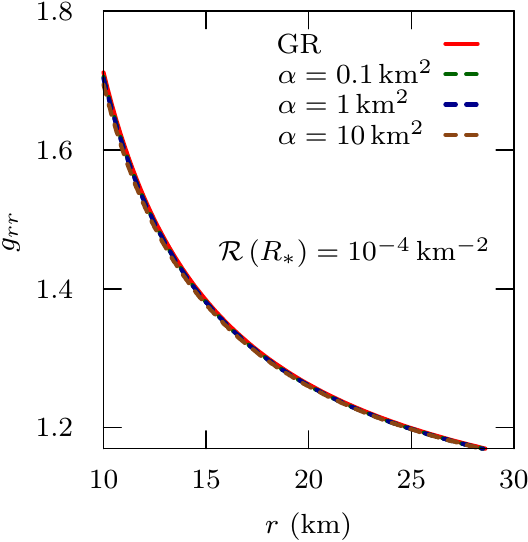}
\\
\includegraphics{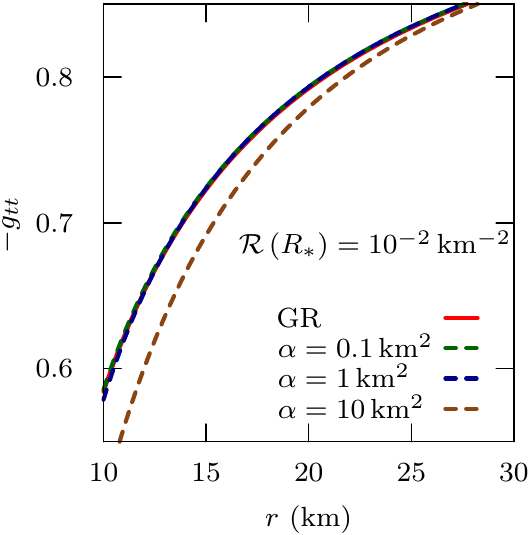}
\includegraphics{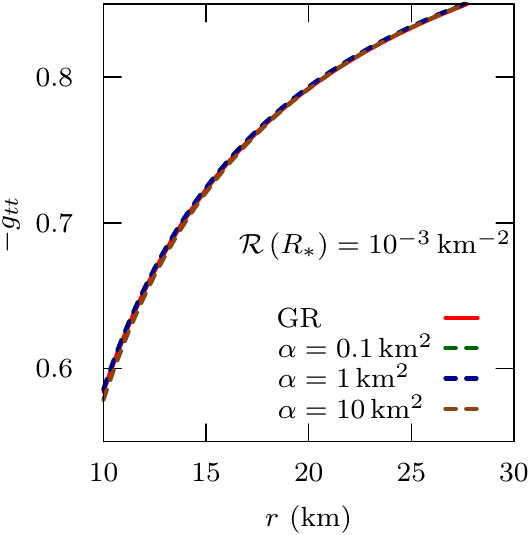}
\includegraphics{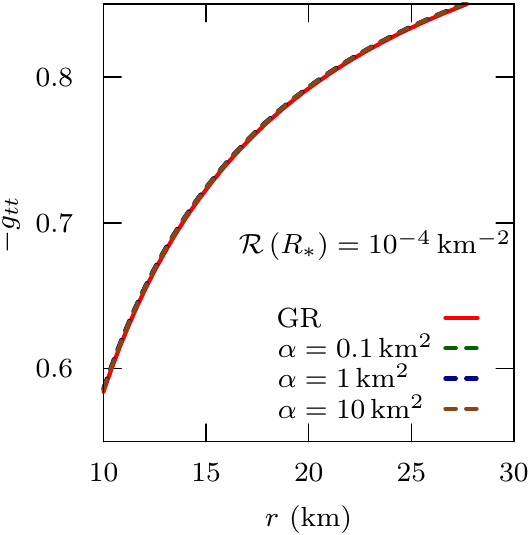}
\\
\includegraphics{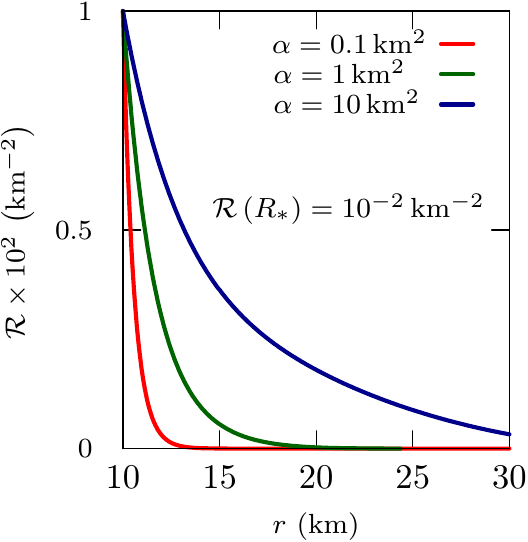}
\includegraphics{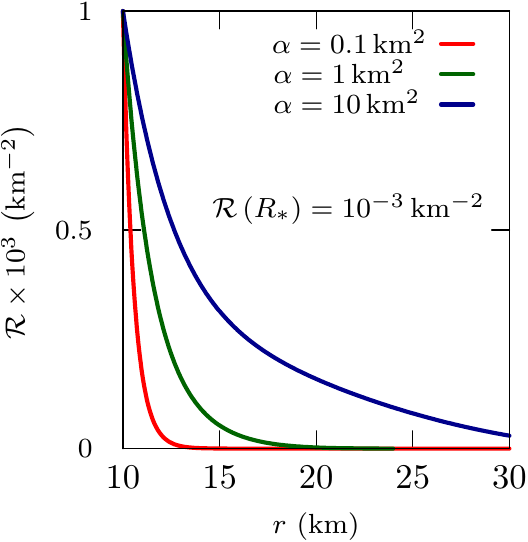}
\includegraphics{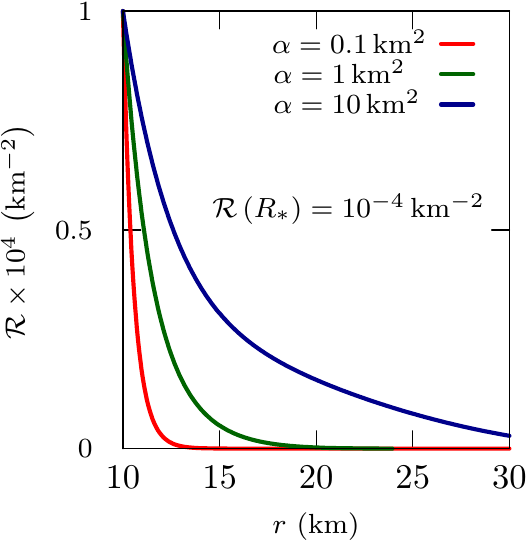}
\caption{The metric components and the Ricci scalar for different values of $\alpha$ and $\mathcal{R}_*$. In all graphs, 
the mass of the star is assumed $1.4\,M_{\odot}$ where $M_{\odot}$ is the solar mass,
and the radius of the star is assumed $R_*=10\,{\rm km}$. \label{fig1}}
\end{figure*}

The metric components and the Ricci scalar for arbitrary values of $\alpha$ and the Ricci scalar at the surface of the star, $\mathcal{R}_*$, are shown  in Fig.\ \ref{fig1} which shows that
the solutions converge to the Schwarzschild's solution far from the star.
The Ricci scalar goes to zero more rapidly as $\alpha$ decreases.
The value of $\mathcal{R}_*$ does not change the
form of the solutions of the Ricci scalar; it makes a difference only at
the magnitude of them. Similarly, the metric components converge to the Schwarzschild's
solution more rapidly as $\alpha$ decreases. Also, the difference between metric components in the Starobinsky model and in GR lessens
as $\alpha$ and $\mathcal{R}_*$ decrease. In the Starobinsky model,
the metric components are smaller than their values in GR and
the value of $g_{tt}g_{rr}$ is less than 1
near the surface of the star unless $\mathcal{R}_*$ is not zero.

The singular points of the metric components are $r=0$ and $r=2M=R_{\rm sch}$ as is the case in Schwarzschild's solution. So, our solutions do not alter the location of the event horizon. The authors of Ref.~\cite{Sal+16} showed that the Ricci scalar should be zero at the Schwarzschild radius in the Starobinsky model for the spherically symmetric and static configuration. So, the outer solutions are valid from infinity to the event horizon, and the solutions given in Eqs.\ \eqref{f_vac}-\eqref{R_vac} reduce to Schwarzschild's solution. Then, Schwarzschild's solution describes the vacuum solution around a static black hole in Starobinsky model when the second term Eq.\ \eqref{action} is assumed to be a perturbative correction to GR.

\section{Conclusion \label{sec:Conc}}

In this paper, we solved the vacuum field equations around
a spherically symmetric and static object in the Starobinsky model by using the MAE method.
We showed that
the boundary layer can occur only near the surface of the star.
Inside the boundary layer the solutions can be different from Schwarzschild's solution
while the metric definitely matches the Schwarzschild solution outside the boundary layer.

We thus conclude that vacuum solutions are not unique in the Starobinsky model. Our solutions have arbitrary
constants which can be determined depending on the value of the Ricci scalar 
at the surface of the star. For some cases, the metric can be different from 
the Schwarzschild metric near the surface of the star. 
With the specific choice that the Ricci scalar is zero at the surface of the star, a boundary layer
does not occur and the Schwarzschild metric is valid all over the vacuum. 
Another case for which a boundary layer does not occur is the  
choice of the boundary condition such that the first derivative
of the Ricci scalar is zero at the surface of the star.
Accordingly, the radius of the star and the value of the Ricci scalar at the surface as well as the mass of the star are required to uniquely define the metric outside spherically symmetric and static stars in the Starobinsky model
while  only mass is required in GR. The Ricci scalar is a geometric parameter rather than an observable or measurable
parameter like the mass and the radius. So, it is reasonable to expect that there is a unique value of the Ricci scalar for all objects 
or it depends on another parameter of the object such as its compactness. Showing this is beyond the scope of this paper.

The vacuum solutions found by Ref.~\cite{Coo+10} reduce to  Schwarzschild's solution in the absence of the cosmological constant. The regular perturbation approach the authors have employed  only provide solutions for the metric outside the boundary layer, and so they miss the solutions other than Schwarzschild's solutions.
This is yet another example which shows the requirement
of using the MAE method when the second term in the Lagrangian is assumed to be
perturbative. 

Our vacuum solutions are different from the solutions obtained in
\cite{Res+16} for the Starobinsky model as well. We considered $\alpha$ to be positive to avoid ``ghosts'' while their solutions are valid only when $\alpha$ is negative.
Still, we could obtain such type of solutions, reincreasing the  absolute value of the Ricci scalar,
by proposing the boundary layer at somewhere other than the surface of the star. Yet, this case
corresponds to increase of gravity with radial distance as described in Appendix \ref{app:C2}.  We, thus, avoided this case purposely.

Our solution of the Ricci scalar does not contain logarithmic or some powers of $r$
as in Ref.\cite{Pun+08}. The difference probably arises due to the different approaches employed
or the difference between the generality of the solutions.

Within the framework of this paper, we found that the vacuum solution around a static black hole is Schwarzschild's solution as consistent with previous studies \cite{Mayo+96,Psa+08,Sot+12,Sal+16,Mig+92}.

In the Starobinsky model, the interior solutions of relativistic stars
also depend on the value of the Ricci scalar \cite{Arap+16}
as well as the vacuum solutions.
The problem being non-well posed prevents finding unique solutions.
The metric components being different from that of Schwarzschild's solution modifies the definition of the mass of the star.
These issues require the simultaneous solution of the interior and the exterior metrics of the star.
Therefore, the self-consistent approach which is employed in \cite{Yaz+14a,Yaz+14b,Yaz+15a}, can be considered as an 
appropriate method to study the structure of relativistic stars in the Starobinsky model. Still, the different boundary conditions cannot be distinguished 
far from the star, and they all satisfy the asymptotic flat space-time as shown in Fig. \ref{fig1}. 
This degeneracy reduces the reliability of using the shooting-method unless the computer code seeks for all possible solutions and does not terminate once a solution is found. 
Continuity of the derivative of the Ricci scalar condition at the surface of the star, given in Eq.\ \eqref{Rprim_con}, can be used to eliminate the solutions 
and might give a unique solution.
Furthermore, it is well expected that different boundary conditions caused by vacuum solutions give different mass-radius relations for neutron stars.

\begin{acknowledgments}
The author thanks K.~Yavuz Ek\c{s}i for encouragement and useful discussions. Also, the author thanks the referee for comments that helped to improve the manuscript.
\end{acknowledgments}

\appendix
\section{Boundary Layer at the Farthest Point \label{app:C2}}
If we find the inner solutions for the boundary layer that occurred
at the farthest point of a fictitious region, then follow the same steps in Sec.\ \ref{sec:VS}, we would obtain the vacuum solutions of the
metric components as
\begin{widetext}
\begin{align}
f^{\rm comp}\left(r\right)=& 
\left[1-\frac{2M}{r}\right]^{-1}
+
\sqrt{6\alpha C_0^3}\frac{K_0R_+}{3}
\exp\left(-\sqrt{\frac{C_0}{6\alpha}}\left(R_+-r\right) \right)
+\alpha\frac{5R_+^2C_0^2K_0^2}{6}
\exp\left(-2\sqrt{\frac{C_0}{6\alpha}}\left(R_+-r\right) \right)
\notag \\
&
+\frac{C_0^2K_0}{12}\left( 1-C_0\right)\left(R_+-r\right)^2
\exp\left(-\sqrt{\frac{C_0}{6\alpha}}\left(R_+-r\right) \right)
-\sqrt{\alpha}\frac{7K_0\sqrt{6C_0^3}}{12}\left(1-C_0\right)
\left(R_+-r\right)
\exp\left(-\sqrt{\frac{C_0}{6\alpha}}\left(R_+-r\right) \right)
\notag \\
&
+\alpha\left[ \frac{C_0K_0}{4}
\left(1-5C_0\right)+\frac{K_1C_0R_+}{3}
\sqrt{6C_0}\right]
\exp\left(-\sqrt{\frac{C_0}{6\alpha}}\left(R_+-r\right) \right),
\\
h^{\rm comp}\left(r\right)=& 
1-\frac{2M}{r}+
\alpha \frac{K_0}{C_0}\left(C_0-3\right)
\exp\left(-\sqrt{\frac{C_0}{6\alpha}}\left(R_+-r\right) \right),
\end{align}
where
\begin{equation}
K_0=\frac{H_0}{L^2}, \qquad K_1=\frac{H_1}{L^3}
\end{equation}
and they are determined according to the value of the Ricci 
scalar at  the farthest point of the region. The behaviors of the 
metric components are shown in Fig.\ \ref{fig_c2} with  two 
different boundary conditions 
where the Ricci scalar has the same magnitude but opposite sign at the farthest point of the region.  It can be seen
from the graphs that the metric 
components are independent of the sign of the Ricci scalar,
and they diverge from $1$ as getting far away from the star.
It means that the gravitation increases as getting far away from the star, and this is not physically reasonable.
\begin{figure*}[h!]
\center
\includegraphics{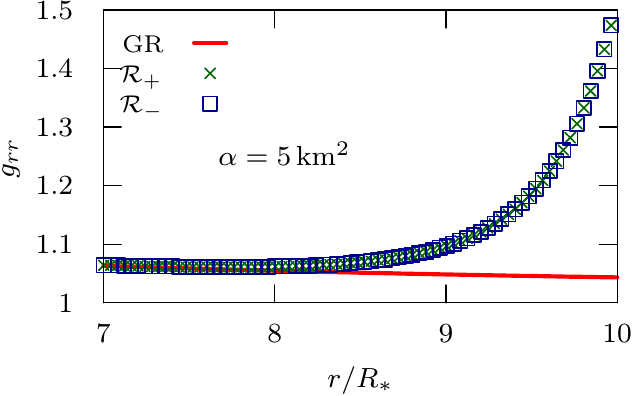}
\hspace{0.5cm}
\includegraphics{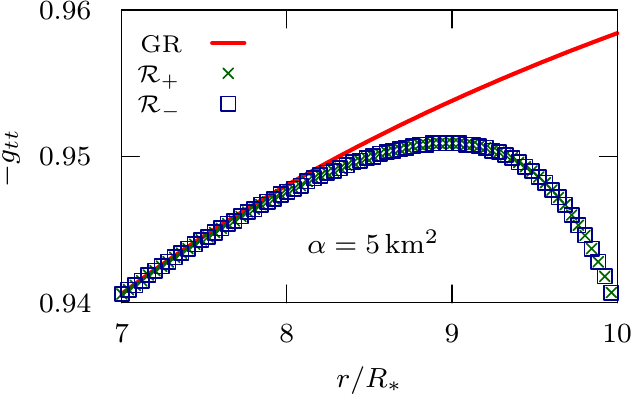}
\caption{The metric components are shown in the case of the boundary layer occurred at the farthest point of a region.
Here, the mass and the radius of the star are taken as 
$1.4\,M_{\odot}$ and $R_*=10\,{\rm km}$, respectively, where
 $M_{\odot}$ is the solar mass. $\mathcal{R}_+$ corresponds to the case where the boundary condition is chosen as positive finite value for the Ricci scalar at $r=10\,R_*$,
and $\mathcal{R}_-$ corresponds to the case where the Ricci scalar at $r=10\,R_*$ has the same magnitude but opposite sign with the $\mathcal{R}_+$ case. \label{fig_c2}}
\end{figure*}
\end{widetext}

\bibliography{ref}
\bibliographystyle{apsrev4-1}

\end{document}